\documentclass[sigconf]{acmart}

\usepackage{subcaption}
\usepackage{multirow}
\usepackage{natbib}
\usepackage{makecell}
\usepackage{xcolor}
\usepackage{setspace}
\usepackage{balance}
\usepackage{hyperref}

\AtBeginDocument{%
  }

\setcopyright{none}
\copyrightyear{2025}
\acmYear{2025}
\acmConference[AutomotiveUI '25]{17th International Conference on Automotive
User Interfaces and Interactive Vehicular Applications}{September 21--25,
2025}{Brisbane, QLD, Australia}

\acmBooktitle{17th International Conference on Automotive User Interfaces and
Interactive Vehicular Applications (AutomotiveUI '25), September 21--25, 2025,
Brisbane, QLD, Australia}
\acmDOI{10.1145/3744333.3747814}
\acmISBN{979-8-4007-2013-0/2025/09}

\acmSubmissionID{1089}

\begin{document}

\title[Socially Adaptive Autonomous Vehicles] {Socially Adaptive Autonomous Vehicles:\\Effects of Contingent Driving Behavior on Drivers' Experiences}

\author{Chishang ``Mario'' Yang}
\affiliation{%
  \institution{Cornell University}
  \city{New York}
  \country{USA}}
\email{cy546@cornell.edu}

\author{Xiang Chang}
\affiliation{%
  \institution{Cornell University}
  \city{New York}
  \country{USA}}
\email{xc529@cornell.edu}

\author{Debargha Dey}
\affiliation{%
  \institution{TU Eindhoven}
  \city{Eindhoven}
  \country{The Netherlands}}
\email{d.dey@tue.nl}

\author{Avi Parush}
\affiliation{%
  \institution{Israel Institute of Technology}
  \city{Haifa}
  \country{Israel}}
\email{aparush@technion.ac.il}

\author{Wendy Ju}
\affiliation{%
  \institution{Cornell University}
  \city{New York}
  \country{USA}}
\email{wendyju@cornell.edu}

\renewcommand{\shortauthors}{Yang et al.}

\begin{abstract}
Social scientists have argued that autonomous vehicles (AVs) need to act as effective social agents; they have to respond implicitly to other drivers' behaviors as human drivers would. In this paper, we investigate how contingent driving behavior in AVs influences human drivers' experiences. We compared three algorithmic driving models: one trained on human driving data that responds to interactions (a familiar contingent behavior) and two artificial models that intend to either always-yield or never-yield regardless of how the interaction unfolds (non-contingent behaviors). Results show a statistically significant relationship between familiar contingent behavior and positive driver experiences, reducing stress while promoting the decisive interactions that mitigate driver hesitance. The direct relationship between familiar contingency and positive experience indicates that AVs should incorporate socially familiar driving patterns through contextually-adaptive algorithms to improve the chances of successful deployment and acceptance in mixed human-AV traffic environments.

\end{abstract}



\begin{teaserfigure}
  \includegraphics[width=\textwidth]{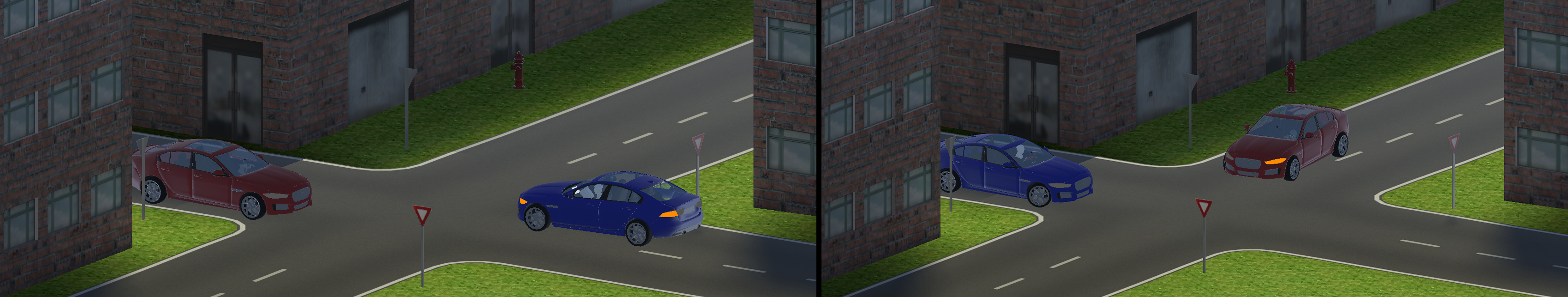}
  \caption{We conducted a Virtual Reality driving simulation study to explore interactions between a human driver and an autonomous vehicle at traffic intersections in a pseudo-naturalistic setting to understand the influence of contingent driving behaviors in different contexts (two of four scenarios shown above).}
  \Description{The image shows two samples of traffic intersection near brick apartment buildings, where in the left frame a red car is waiting at a yield sign while a blue car starts to pass. In the right frame, the blue car waits at the intersection while the red car starts to proceed to turn.}
  \label{fig:teaser}
\end{teaserfigure}

\maketitle

\section{Introduction}
Traffic interactions are fundamentally social activities where drivers must coordinate their actions through implicit communication. As Swan and Owen describe, driving interaction is ``the presentation of action, a request for reaction,'' where drivers are ``embedded in a social situation created by the actions of others.'' ~\cite{swan1988social} This social dimension presents a significant challenge for autonomous vehicles (AVs), which need to navigate not only physical environments but also social landscapes with complex, often unwritten rules.

Prior research has emphasized the importance of implicit communication in traffic interactions, particularly through vehicle movement and kinematics. However, despite this recognition of driving's social nature, we have not found prior research that has explored in a controlled study how reciprocal interaction from AVs affects human drivers.

In this project, we designed a contextually-adaptive algorithm based on data from human driving interactions in ambiguous yielding scenarios at intersections. Our model responds to other drivers' behaviors in ways that match familiar social norms, unlike conventional AV approaches that follow rigid rules regardless of context. We tested this responsive model against two non-contingent alternatives (always-yield or never-yield) in a controlled virtual reality driving simulator study to measure how familiar contingent behavior affects drivers' experiences. The statistically significant results indicate that AVs exhibiting familiar contingent behavior create more positive interaction experiences for human drivers.

Our findings suggest that autonomous vehicles should incorporate algorithms that respond to other drivers in familiar ways. This approach could improve AV acceptance and integration into existing traffic environments by aligning with drivers' expectations based on established social interaction patterns on the road.

\subsubsection*{Contribution statement}
This paper explores how adaptive movement from autonomous vehicles (AVs) influences the behavior and experiences of other road users. Specifically, we show that contingent driving interactions from AVs can reduce stress and hesitance for human drivers compared to non-contingent interactions that simply yield or take the lead regardless of context. These findings highlight the importance of considering other vehicles' implicit behavior when developing AV algorithms to seamlessly integrate AVs into traffic as socially aware agents.
\section{Background}
\subsection{Interaction in Driving}
Swan and Owen have described driving interaction as ``the presentation of action, a request for reaction,'' and noted that ``the very nature of driving behavior dictates that the driving public are embedded in a social situation created by the actions of others.'' ~\cite{swan1988social} In this way, a communication model might offer the best analog for driving interaction. Linguist Herb Clark describes the way that people coordinate with each other in a material world as ``material signals''---``signals in which they indicate things by deploying material objects, locations, or actions around them.'' ~\cite{clark2005coordinating} With this lens, the explicit verbal or gestural signals we use to communicate with one another on the street are akin to the implicit signals we exhibit with our bodies or through our vehicles---both outgrowths of social interaction that we use to negotiate joint action ~\cite{winograd1986understanding}. In a 1992 article in \textit{Accident Analysis and Prevention}, David Zaidel posited the following possible mechanisms by which ``others'' can affect the behavior of individual drivers: others as a source of information, communication with others, others as a reference group, imitation of others ~\cite{zaidel1992modeling}. So the literature highlights how driving is a \textit{contingent} activity.

Here, we distinguish `contingent driving behavior' as being distinct from merely `socially-aware driving'; socially aware driving involves context-sensitive responses to environmental and social cues such as interpreting gestures or yielding to pedestrians, but contingent behavior addresses the broader behavioral principles from experimental psychology describing interactants ``responses based on the consequences of the preceding actions ~\cite{10.1037/h0045185}.'' Whereas socially-aware driving might describe, for example, the way a driver responds to social factors in the environment, such as pedestrian density, or the presence of older adults, contingent behavior would describe the way one driver responds behaviorally to the specific preceding behaviors of other pedestrians or drivers.

One potential challenge of studying driving interaction is that typical interactions might vary by region. Sociologist Dale Dannefer, for example, mentions informal norms such as following distance, merging behavior, right-of-way rules, but also performance of attention or inattention ~\cite{dannefer1977driving}. As drivers ourselves, we have experience with the challenges of driving in unfamiliar places; even if we know how to operate the car, we do not understand the ``lingua franca'' of the road. Sociologists therefore note that some crashes are not the result of individually risky behaviors but rather the results of ``social accidents,'' caused by interactions between people from different social groups interpreting and responding to situations differently ~\cite{factor2007social}. These differences are difficult to study because many of the key variations in interaction are implicit ~\cite{dong2024XCdifferences}. Still, we believe it is important for AVs to understand and adopt local driving interactive response, which means the driving is ideally not only contingent, but \textit{normal}--that is, familiar to the human drivers in the setting.

\subsection{Using Driving Simulation\\to Understand Interaction}
One of the challenges to developing normal contingent driving has been the question of how to capture driving interactions. Until recently, most of the research on driving interaction was based on direct or recorded observation. Juhlin, for example, employed ethnographic techniques observing students at a Swedish driving school, interviewing participants, recording driving sessions and transcribing and thematically coding incidents of cooperation between road users ~\cite{juhlin1999traffic}. Similar investigations have been made of social agent navigation in urban traffic ~\cite{toiskallio2002impersonal}, driver-bus interaction ~\cite{normark2006enacting}, pedestrian-vehicle interaction ~\cite{vsucha2014road, dey2017pedestrian} and interactions at petrol stations ~\cite{normark2006tending}. Vinkhuyzen and Cefkin used ethnographic techniques to understand how autonomous vehicles will engage with pedestrians, bicyclists and other cars in a socially acceptable manner, and noted the difficulty of making observational distinctions with these methods ~\cite{vinkhuyzen2016developing}.

Virtual reality is increasingly used by researchers seeking to better understand driving interaction. Recent research has used virtual simulations of autonomous vehicles to understand how pedestrian interactions with autonomous vehicles are influenced by vehicle appearance and behavior ~\cite{dey_understanding_2001, burns2019pedestrian}, and the utility of external displays to indicate vehicle intent ~\cite{hollander2019investigating, li2018cross}.

The advent of multi-driver simulation platforms has opened up possibilities for driver interaction studies. The use of multi-driver simulation studies to examine the interaction between drivers was first performed by Hancock and De Ridder in 2003 ~\cite{hancock2003behavioural} who placed two participants into adjacent full-vehicle simulators that share a single virtual world to understand collision avoidance behaviors. In 2011, Muhlbacher et al. developed a platform to study interactions between four drivers in a platooning scenario ~\cite{muhlbacher2011multi}. Researchers at the Institute for Transportation Studies at the German Aerospace Center (DLR) created a Modular and Scalable Application Platform for ITS Components (MoSAIC) to understand interactions between V2V connected vehicles and non-equipped vehicles ~\cite{oeltze2015benefits}, cooperative lane change maneuvers ~\cite{heesen2012investigation}, and traffic-light assistance ~\cite{rittger2015anger}. More recently, however, the automotive UI community has taken on driving interactions as a major research area ~\cite{markkula2020defining, kalantari2023testing, goedicke2022strangers, goedicke2023driving, bazilinskyy2024not, dey_2025_operationlize}. Simulators make it possible to better control and instrument the driving environment, so as to focus on the effects of multi-party interactions. Naturalistic driving studies, such as ~\citet{deligianni2017analyzing,victor2015analysis} in contrast, very often lack key information about the driving context that is needed to understand the driving behavior. Of course, there are ecological validity concerns with the use of driving simulations, even highly immersive first-person virtual reality driving simulators such as the one used for this research study. Nevertheless, even if differences between real on-road driving and VR simulation driving exist, the controlled environment of the driving simulator makes it possible to study the \textit{relative} effects of interactive driving behaviors which ethnographic and naturalistic studies have established are important to understand.

\subsection{Algorithmic Driving}
While Zaidel posited the possibility of formalizing the interactive model between drivers as a mathematical model that would enable prediction of behavioral mixes in 1992, suggesting that computer and laboratory simulation would be useful methods for beginning the research, actual simulator studies of driver interaction are a recent phenomenon ~\cite{zaidel1992modeling}.

The advent of automation has also made it possible to study interaction by allowing programmatic control over one of the interacting agents to understand how differences in the AV behavior influences drivers. 

Sadigh et al., for example, have created simple driving simulation studies which feature an interaction between an autonomous car and a human driver modelled as a dynamical system, in which the robot's actions have immediate consequences on the state of the car, as well as on human actions. They model these consequences by approximating the human as an optimal planner, with a reward function that is acquired through Inverse Reinforcement Learning, and performed user studies to verify that the robots can indeed influence human actions ~\cite{sadigh2016planning}.

The prior literature, however, indicates that driving should not only be communicative but familiar. Past research highlights the importance of implicit cues such as motion patterns which other road users use to decipher the intention of the AV~\cite{dietrich_implicit_2020,rettenmaier_communication_2021,Moore2019ExplicitEhmi}. To develop an algorithmic driving model that incorporates contingent norms, the AV needs not only to understand how to communicate or motivate human drivers, but also to be able to respond in an expected manner to the human drivers' own behaviors. This suggests that data captured about driver-driver interactions would have to be incorporated into models to mimic the contingent and normal interactions drivers expect. For this research study, we used data captured from prior research in driver-driver interactions to build dynamic algorithmic models of AV driving that use the human driving behavior as a key input~\cite{XCDrivingData}.
\section{Research Question}
Our goal is to investigate the role of familiar contingent behavior in driver-AV interactions. Familiar contingent behavior refers to the ability to dynamically respond to another driver's actions in accordance with expected social norms, as opposed to following rigid, predetermined patterns irrespective of how the interaction unfolds. This leads us to the following research question:
\begin{quote}
\textit{How does a familiar contingent driving behavior impact a human driver's experiences interacting with AVs?}
\end{quote}

We hypothesize that drivers not only expect contingency from one another, but also that the contingency is dealt with in a way that is in alignment with familiar social norms, and therefore non-contingent behavior will cause hesitance and stress. Our sub-hypotheses are thus:
\begin{itemize}
    \item $H_1$: A familiar contingent AV driving behavior will elicit less hesitance than a non-contingent behavior
    \item $H_2$: A familiar contingent AV driving behavior will lead to a more relaxed interaction than non-contingent behavior
    \item $H_3$: A familiar contingent AV driving behavior will cause less stress than non-contingent behavior
\end{itemize}

\section{Method}
We evaluated our research question in a mixed-design Virtual Reality (VR) experiment, which was approved by the Internal Review Board of the researchers' institution(s) under IRB \#0147944.

\subsubsection*{\textbf{Participants}}
We recruited 56 participants for the study from New York City residents who hold US driver's licenses, through a variety of channels including university databases, social media, and public posters. Due to motion sickness, 6 participants stopped partway through the study and were excluded from data analysis, resulting in 50 participants who completed the study (24 males, 25 females, 1 non-binary; mean age 27.48 years; SD = 8.30).

\subsubsection*{\textbf{Task}}
Participants were instructed to drive in a virtual reality traffic environment using a steering wheel, pedals, and control interfaces while wearing a VR headset. They were asked to drive as they would in the real world while following turn-by-turn navigation instructions displayed on a virtual screen. Each participant experienced 12 trials (4 scenarios $*$ 3 different AV behaviors) where they encountered and interacted with different AV behaviors at intersections.

\subsection{Study Design}
\label{StudyDesign}

\begin{figure*}[htbp]
    \centering
    \begin{subfigure}[b]{0.25\textwidth}
        \centering
        \includegraphics[width=\textwidth]{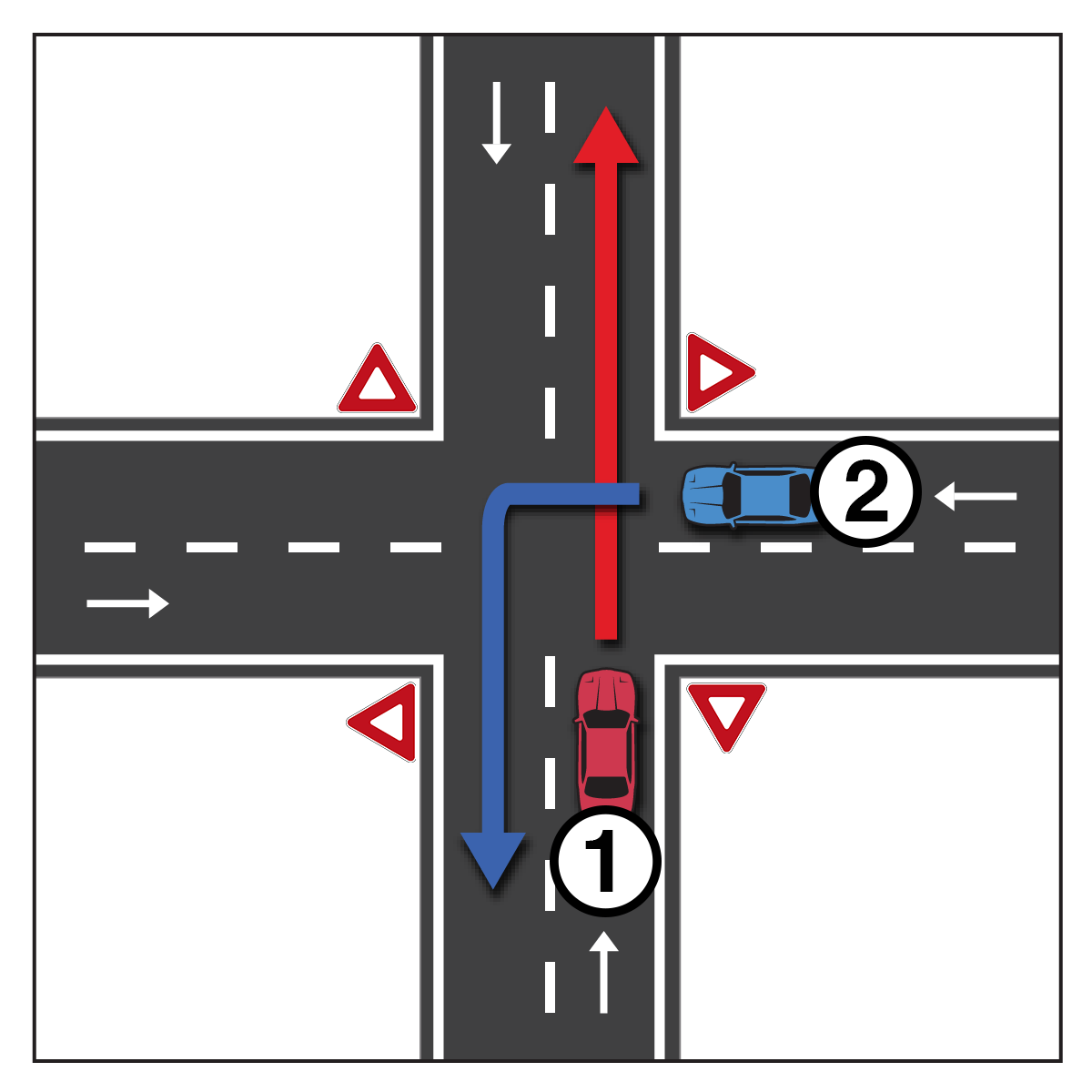}
        \caption{S1: Orthogonal approach, mixed}
        \Description{Vehicle 1 from north goes straight while Vehicle 2 from east turns left}
    \end{subfigure}%
    \begin{subfigure}[b]{0.25\textwidth}
        \centering
        \includegraphics[width=\textwidth]{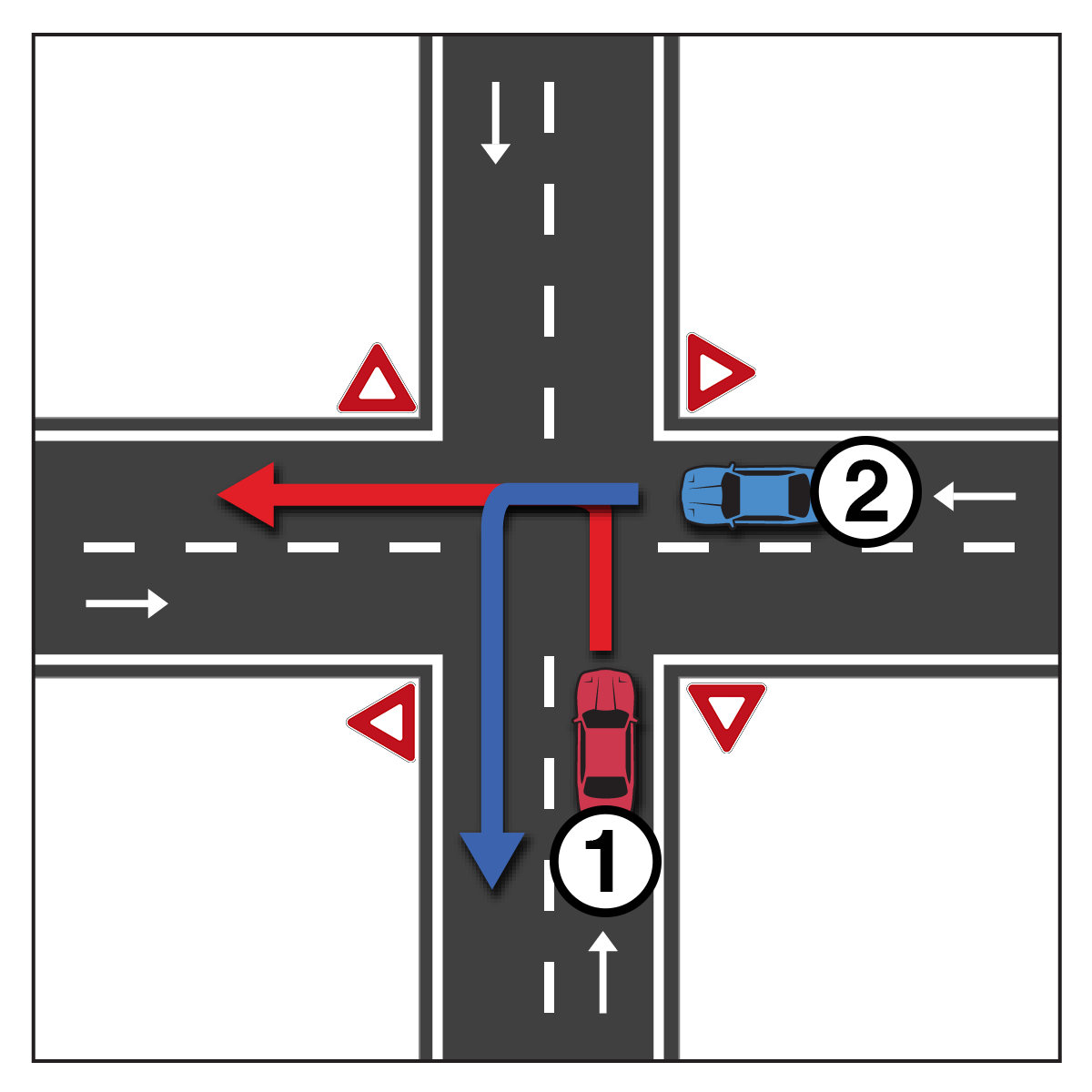}
        \caption{S2: Orthogonal, turning left}
        \label{ScenarioS2}
        \Description{Vehicle 1 from west goes left while Vehicle 2 from east turns left}
    \end{subfigure}%
    \begin{subfigure}[b]{0.25\textwidth}
        \centering
        \includegraphics[width=\textwidth]{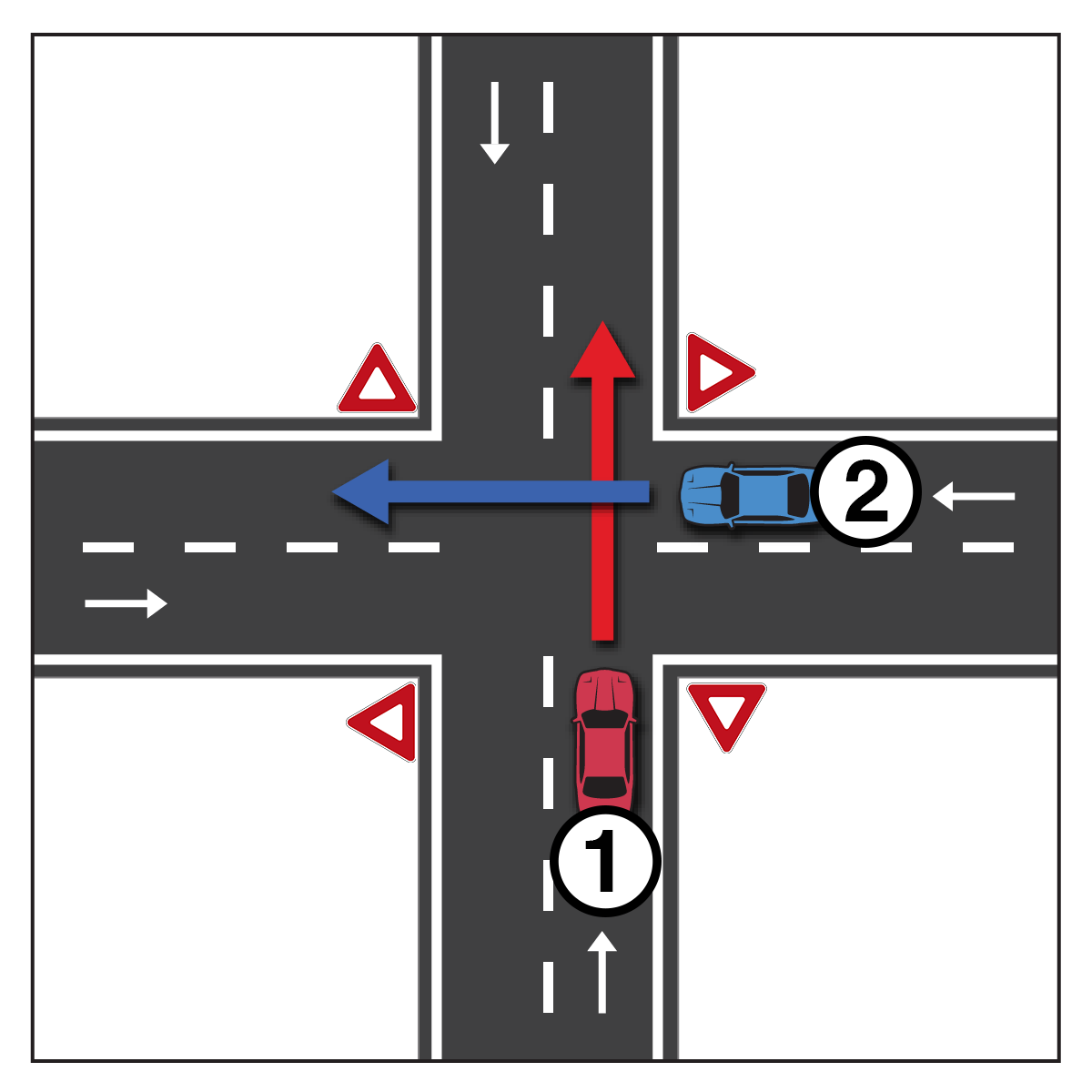}
        \caption{S3: Orthogonal, going straight}
        \Description{Vehicle 1 from north goes straight while Vehicle 2 from west goes straight}
    \end{subfigure}%
    \begin{subfigure}[b]{0.25\textwidth}
        \centering
        \includegraphics[width=\textwidth]{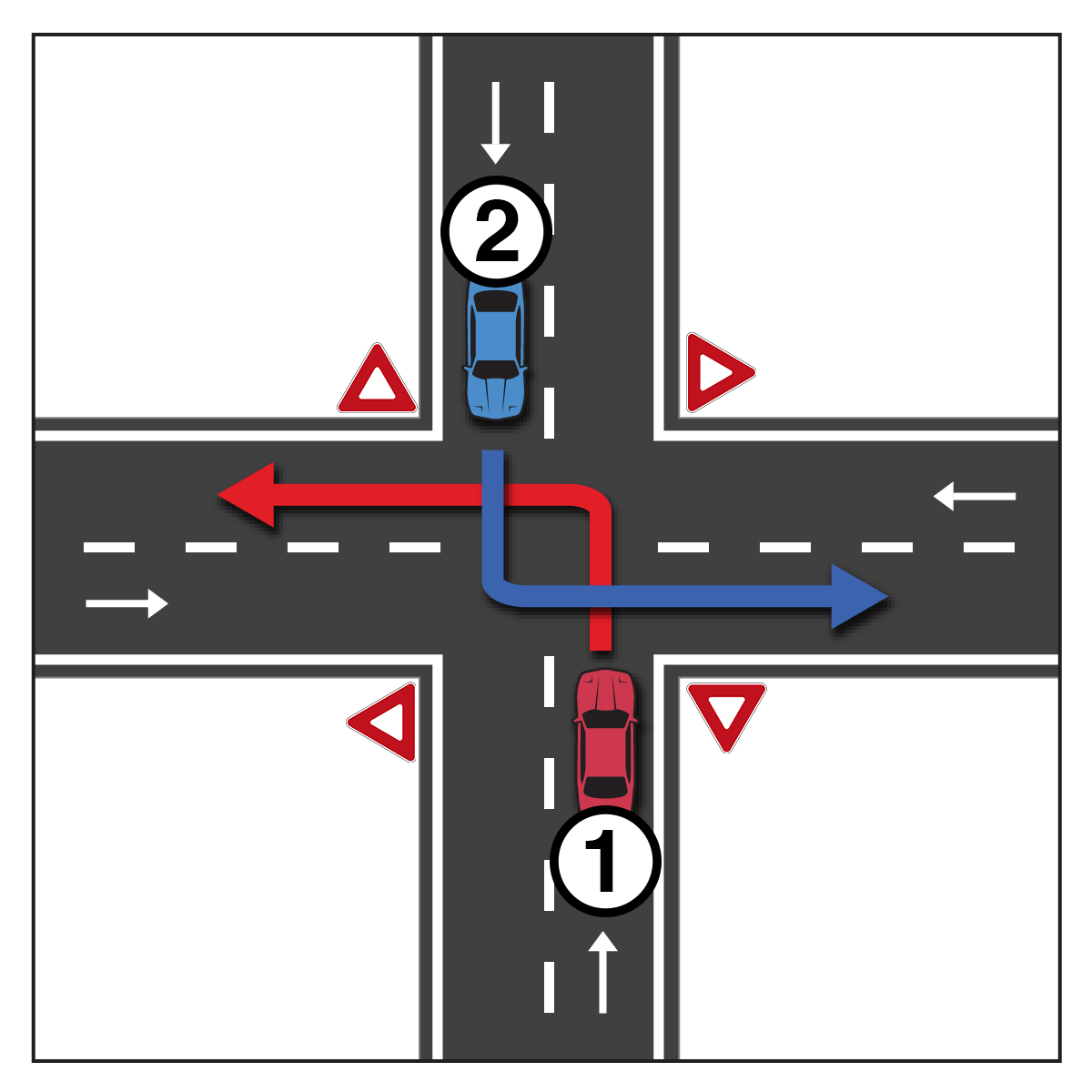}
        \caption{S4: Opposing, turning left}
        \Description{Vehicle 1 from south turns left while Vehicle 2 from north turns left}
    \end{subfigure}
    \caption{The four scenarios selected for study}
    \label{fig:all_CPs}
\end{figure*}

In this study, we used four distinct traffic interaction scenarios to investigate our research question. The interaction scenarios were grounded in the full factorial design of a dyadic vehicle interaction at a traffic intersection as outlined in the work of Dey et al.~\cite{dey_2025_operationlize}. Of the 27 scenarios that emerged from their full-factorial design, we conducted pilot tests to identify the ones that resulted in the highest number of space sharing conflicts~\cite{markkula2020defining}. This resulted in four scenarios that serve as the control variables for our study, as shown in Figure~\ref{fig:all_CPs}.

Specifically, we chose the setup of unsignalized intersections so that the interaction is not dictated by traffic lights, but rather by drivers exercising judgment and negotiating based on the position and behavior of the cars. Scenario S1 involves two cars approaching orthogonally, where the car approaching from the relative right (car~2) makes a left turn, and the other car (car~1) goes straight. S2 is similar to S1, as the cars also approach orthogonally, but in this scenario, both cars make a left turn. In S3, the cars once again approach each other orthogonally, and both cars go straight. Finally, in S4, the cars approach from opposite directions, and both make a left turn. In each of these scenarios, we placed yield signs on both approaching roads, as there were no traffic lights. Our rationale was to create scenarios in which the cars would need to slow down and consider the situation and the other approaching vehicle before making a decision about entering the intersection.

\subsubsection{Right-of-way (RoW):}
Right-of-way is defined as ``a precedence in passing accorded to one vehicle over another by custom, decision, or statute''.~\cite{merriam-webster_right--way_2025} In this case, we are specifically interested in RoW for unsignalized equal intersections. Since this study was conducted in the United States, we ground our definition in the North American traffic rules. As an example, the New York Driver's Manual states that ``at intersections not controlled by signs or signals, or where two or more drivers stop at STOP signs at the same time and they are at right angles, the driver on the left must yield the RoW to the driver on the right.''~\cite{new_york_department_of_motor_vehicles_ny_dmv_new_nodate} Prior research indicates that drivers in different locales might be more or less attuned to RoW versus order of arrival for determining which driver should yield in a driving encounter~\cite{klein_modeling_2024, 10919603}. For this study, because RoW is a possible confound for yielding and non-yielding behaviors, we balanced the trials in which drivers have RoW between groups. This means that in our study, in S1, S2, and S3, the blue car (car~2 which approaches from the relative right) has the RoW in the event both cars arrive at the intersection at the same time. The right-of-way in S4 is ambiguous and requires negotiation between drivers.

To ensure a balanced study design, we ensured that both the human driver and the AV were assigned interchangeably to car~1 and car~2 so that we had a mix of setups where the human driver had the right-of-way. To do this systematically, we split the trials in two groups:

\begin{itemize}
    \item RoW Group A: AV has the right-of-way [i.e. AV is assigned to the blue car (car~2) and participant is assigned to the red car (car~1)]
    \item RoW Group B: Participant has right-of-way [i.e. AV is assigned to the red car (car~1) and participant is assigned to the blue car (car~2)]
\end{itemize}

Table~\ref{tab:participant_sessions} shows the complete session structure for both Group A and B participants, detailing the specific scenarios and AV behaviors presented across all 12 experimental trials.

To promote maximum interaction between vehicles, we designed the AV's driving behavior to ensure it arrived at the intersection at approximately the same time as the human driver so that the right-of-way rule applies. Since scenario S4 was a symmetrical scenario with no clear RoW, we included it in both groups as it represents a symmetrical situation where the geometric positioning of the cars on the road doesn't inherently favor either vehicle. 

\subsubsection{AV Behaviors:}
To answer our RQ, we needed distinct AV behaviors that represented a ``contingent behavior''. To this end, we developed a contextually adaptive algorithm trained on prior experimental data of interactions between human drivers and adapts to how the interaction unfolds to act as a representative of a ``familiar'' driving behavior ~\cite{XCDrivingData}. We contrasted this with behaviors that employed either go or no-go strategies regardless of how the interaction unfolds, representative of non-contingent behavior.

All behaviors employ a finite state machine to inform driving decisions. The contingent model uses past data from driving interactions between two human drivers to derive a decision policy with a neural network model. We employed the Intelligent Driver Model (IDM) as the general framework for the longitudinal movements of all behavior models~\cite{treiber_traffic_2014}. IDM is the basis of adaptive cruise control on free-road and car-following scenarios, which is suitable to serve as the kinematic base model with the neural network model giving the yield/no-yield interactive decisions. Maneuvers across all three behaviors start with matching the human driver's velocity to ensure two vehicles arrive at the intersection at about the same time, and then perform IDM with diverging decision strategies.

\begin{itemize}
    \item \textbf{Contingent Model:} contextually-adapted algorithmic design based on human driving data. This allows the vehicle to make yielding decisions and acceleration maneuvers by consulting the neural network model. If the model determines crossing is inappropriate, the AV continues to decrease speed and keeps consulting the algorithm until conditions become favorable. If crossing is determined to be appropriate, the AV performs IDM through the intersection. 
    
    \item \textbf{Yield:} starts to decelerate, then attempts to maintain a slow and constant speed until reaching the intersection boundary. Once it crosses this boundary, it performs IDM through the intersection. 
    
    \item \textbf{NonYield:} immediately executes IDM, proceeding through the intersection without yielding. 
\end{itemize}

We designed the contingent model to exhibit a familiar behavior, while the other two models were designed to be less familiar. This design is validated in the \hyperref[Results: Familiarity]{results section}. 

In summary, the study setup was a $(2 \times 3 \times 4)$ design around Right-of-Way (2 levels: Human, AV), AV Behavior (3 levels: Contingent, Yield, Non Yield), and Scenarios (4 levels: S1, S2, S3, S4). In order to keep the number of trials per participant manageable (i.e. to keep the total study participation time within 1 hour), we ran the study with the RoW groups as the between-subject variable, and AV behaviors as the within-subject variable. This led to a total number of 12 trials per participant as shown in Table~\ref{tab:participant_sessions}.

\subsection{Apparatus}
\begin{figure*}[htbp]
    \centering
    \begin{subfigure}[b]{.49\textwidth}
        \centering
        \includegraphics[height=15em]{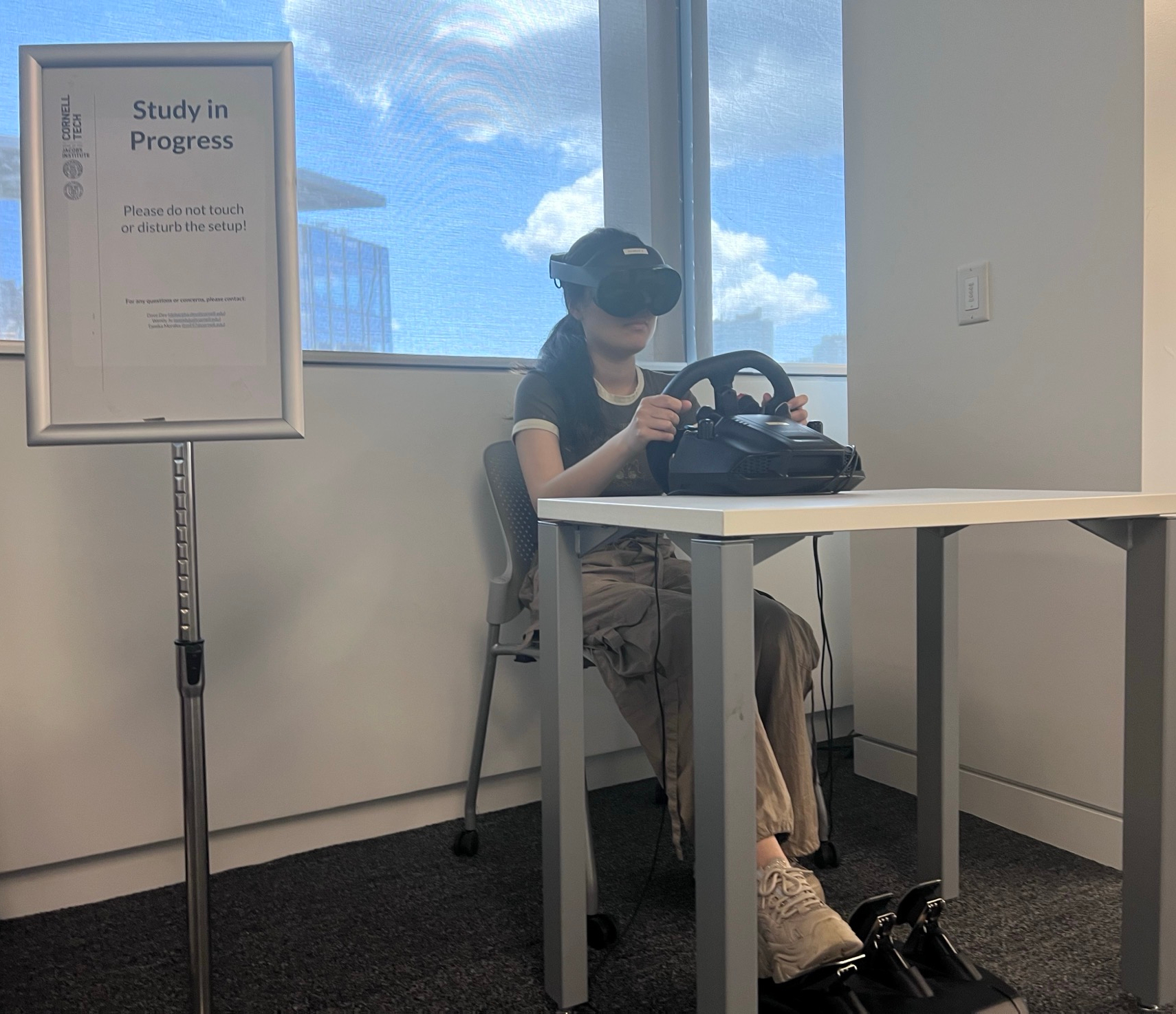}
        \caption{Participant using the simulator with VR headset}
        \Description{A participant sitting at a desk wearing a VR headset while using the driving simulator}
    \end{subfigure}
    \begin{subfigure}[b]{0.49\textwidth}
        \centering
        \includegraphics[height=15em]{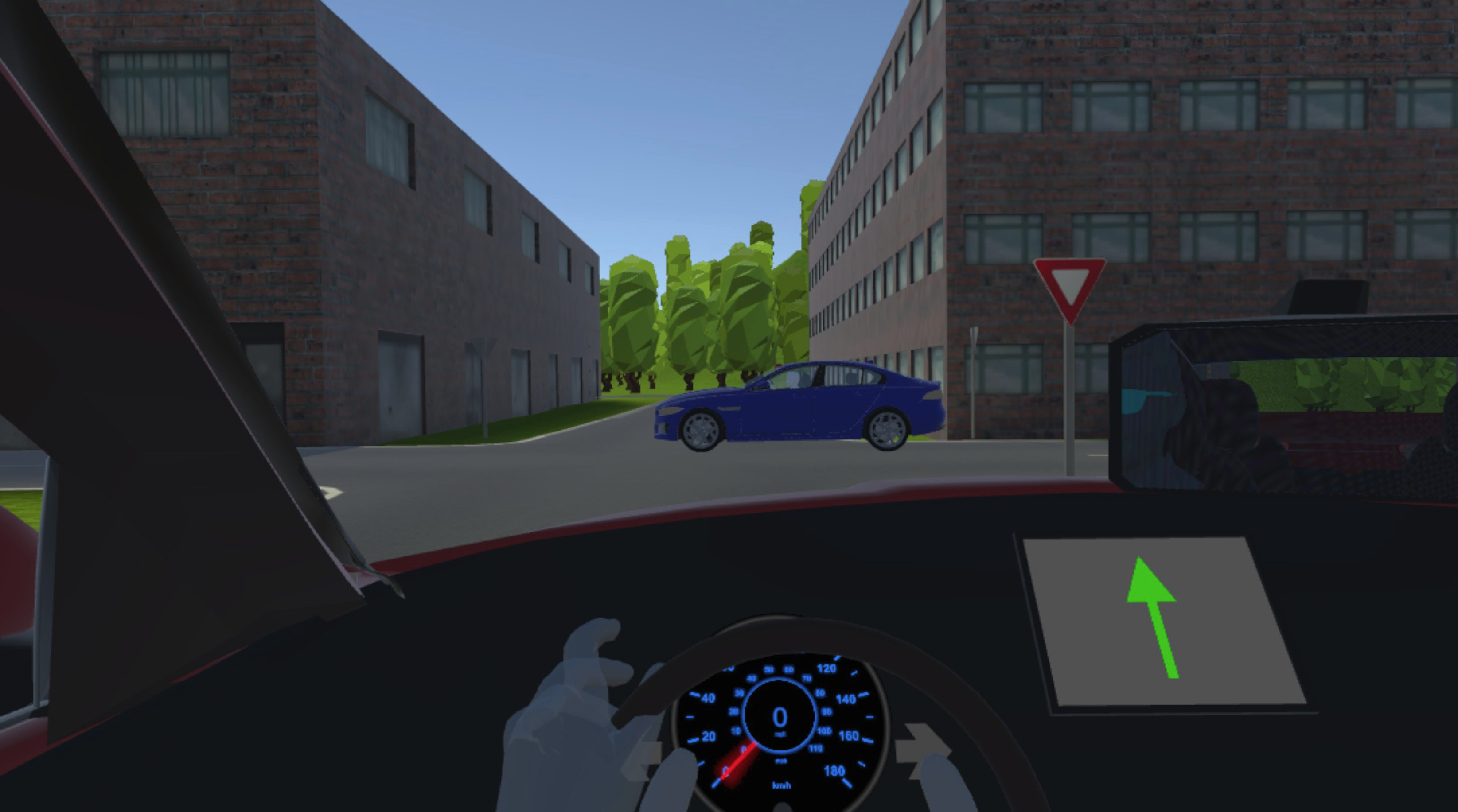}
        \caption{Participant's point of view in VR}
        \Description{Participant's point of view showing a virtual road intersection with buildings and a blue car visible through the windshield}
    \end{subfigure}
    \caption{Experimental setup showing the driving simulator environment}
    \label{fig:simulator-setup}
\end{figure*}

We used Goedicke et al.`s open-source \textit{Strangeland} driving simulator, which enables multi-agent driving interaction in virtual reality~\cite{goedicke2022strangers} to create an immersive experience in a desktop simulation setup. The system uses an Oculus Quest VR headset with a Logitech G29 force-feedback steering wheel to create an immersive driving experience. 

\subsection{Procedure}
The experiment was conducted in a controlled laboratory setting at our research institution. After reviewing and signing the consent form, participants were introduced to the experimental setup. Participants were asked to put on a VR headset and instructed to operate the driving simulator using a steering wheel, pedals, and additional control interfaces (horn, headlights, and turn signals). Driving instructions were displayed on the vehicle's dashboard, and participants were asked to drive as they would in the real world while following turn-by-turn navigation instructions.

The study began with a practice session where participants drove through an intersection without AV's presence. This allowed them to become familiarized with the virtual environment and vehicle maneuvers in VR. Following the practice session, participants were informed that they would encounter AVs in future sessions. They experienced a total of 12 experimental trials, consisting of 4 different scenarios combined with 3 distinct AV behaviors, presented in randomized order. Throughout these scenarios, the system recorded participants' vehicle control inputs, hand movements, and head positions as they negotiated with the AV and navigated through the intersections. After completing each trial, participants came to a complete stop at a ``do not enter'' sign, where they removed the VR headset and answered a questionnaire about their experiences. This process was repeated for all 12 experimental trials.

After concluding all trials, participants completed a demographic questionnaire and participated in a semi-structured interview to reflect on their overall experience with the AV. The interview explored whether certain interactions stood out to them, the major factors they believed determined how interactions with the AV unfolded, and their conception of right-of-way in general. The entire experimental session lasted approximately 50 minutes, and each participant received 15 USD compensation for their participation.

\subsection{Measures}
The study incorporated four different measures. First, we used participants' familiarity with the other vehicle's behavior to ground the AV behaviors with participants' expectations. Subsequently, we collected Likert scale data on self-reported hesitance, relaxation, and stress while interacting with AV to evaluate the effect of familiar driving behavior on human drivers' experiences ~\autoref{tab:questionnaire_items}. These items were developed specifically for this study.

\section{Results}
In this paper, we focus on our insights derived from the participants' self-reported metrics of hesitance, relaxation, stress, and familiarity. Although the simulator recorded a rich set of behavioral data including vehicle motion, head orientation, hand gestures, steering inputs, pedal usage, turn signals, and other driver actions, the detailed analysis of these data is beyond the scope of this paper. Findings based on these data will be reported in a separate publication. 

For the analysis of the self-reported metrics, all statistical tests for main effects are reported at a 0.05 significance level, with a Bonferroni post-hoc adjustment. 
\subsection{Inclusion Criteria}
To ensure valid analysis, we only included trials where two specific criteria were satisfied: (1) an encounter occurred between vehicles, and (2) the human driver consciously noticed the AV.

\textbf{Encounter: } Referring to other studies conducted in the same intersection environment, we also define an encounter using the spatial-temporal requirement of both vehicles being simultaneously present within a range of $5 - 25m$ from the center of the intersection ~\cite{klein_modeling_2024}. We validated that all trials in this study satisfied this requirement.

\textbf{Notice: } To confirm participant awareness, we used the questionnaire item ``As you approached the intersection, did you notice another vehicle?''. Trials were only included when participants explicitly confirmed noticing the other vehicle, ensuring that any measured responses were based on conscious awareness of the AV's presence.

Out of the 600 trials, in 544 trials the participants reported noticing the other vehicle, which will be the basis for analysis.

\subsection{Familiarity of AV Behaviors}
\label{Results: Familiarity}

\begin{table}[htbp]
\centering
\footnotesize
\caption{Group Statistics for Familiarity}
\label{tab: Familiarity_Description}
\begin{tabular}{lcccc}
\hline
\textbf{AV Behavior} & \textbf{Count} & \textbf{Average} & \textbf{Median} &\textbf{Std. Dev.} \\
\hline
Contingent & 188 & 8.27 & 9  &1.47\\
NonYield & 182 & 7.41 & 8  &2.26\\
Yield & 174 & 7.30 & 8  &2.22\\
\hline
\end{tabular}
\end{table}

\begin{table}[htbp]
\footnotesize
\captionof{table}{Kruskal-Wallis Test Results for Familiarity}
\label{tab: Familiarity_ANOVA}
\begin{tabular}{lc}
\hline
\textbf{Measure} & \textbf{Value} \\
\hline
P-Value& $< 0.001$ \\
Test Statistic &21.34\\
$\eta^2$ & $0.036$\\
\hline
\end{tabular}
\hfill
\footnotesize
\captionof{table}{Post-hoc Tests for Familiarity}
\label{tab: Familiarity_Ranked}
\begin{tabular}{llcc}
\hline
\textbf{Behavior 1} & \textbf{Behavior 2} & \textbf{Difference} & \textbf{Adj. P-Value}\\
\hline
Contingent & Yield & 0.97 & <0.001\\
Contingent & NonYield & 0.86 & 0.002\\
NonYield & Yield & 0.11 & 1\\
\hline
\end{tabular}
\end{table}

Our research question hinges upon the supposition that a familiar contingent behavior (i.e., the behavior of an AV that dynamically adapts to interaction around it based on familiar/accepted traffic behavior) has an influence on drivers' experiences. Therefore, we needed to validate that the contingent behavior adequately reflected some level of familiarity for the participants compared to the non-contingent models. To this end, we examined how well the three behaviors represent familiarity by analyzing the self-reported familiarity score, which is shown as a slider from 0 to 10 in the questionnaire. Our RQ rests on establishing the familiarity of the different vehicle behaviors. We are not interested in the other independent variables (Right-of-Way and Scenario) since they are not attributes of the vehicle, and there is no theoretically motivated rationale for their effect on familiarity. Therefore, we aggregated the data across these independent variables and conducted our statistical analysis to establish the effect of Behavior on familiarity. 

Normality and homogeneity of variances were initially assessed for the variable. The results indicated that the normality assumption was satisfied, but failed the homogeneity of variances test. Therefore, we conducted Kruskal-Wallis test with post-hoc pairwise comparisons using Bonferroni correction. Table ~\ref{tab: Familiarity_Description} shows that the Contingent model received the highest familiarity ratings ($Mean = 8.27, Median = 9$) with the most consistent responses ($SD = 1.47$). Both the Non-Yield ($Mean = 7.41, Median = 8, SD = 2.26$) and Yield ($Mean = 7.30, Median = 8, SD = 2.22$) behaviors received lower familiarity scores with greater variability in participant responses. Table ~\ref{tab: Familiarity_ANOVA} revealed a significant overall difference between behaviors ($p < 0.001$).

Post-hoc analysis in Table~\ref{tab: Familiarity_Ranked} shows significant differences between the Contingent model and both non-contingent models ($p = 0.002$ for Non-Yield comparison and $p < 0.001$ for Yield comparison). No significant difference was found between the Non-Yield and Yield behaviors ($p = 1$). This validates our assumption that the behavior generated by the contingent model, trained on real human driving data, is perceived as more familiar to participants than the artificial behaviors of the non-contingent models that always yield or never yield regardless of interaction context.

\subsection{Main and interaction effects of Right-of-Way, Behavior, and Scenario}
A Linear Mixed Model (LMM) was used to examine the effects of Behavior (3 levels: Contingent, Yield, Non Yield), Scenario (4 levels: S1 through S4), and Right-of-way (RoW; 2 levels: Human and AV) on the three participant measures (hesitance, relaxation, and stress) scores. The model included all main effects and interactions as fixed effects, and a random intercept for Participant to account for repeated measures. We discuss the results for each measure separately.

\subsubsection{Hesitance:}
The analysis revealed a significant main effect of Behavior $(F(2,475.5)=6.20,p=.002)$, establishing that the AV’s actions were a key driver of participants' hesitance. However, the effect of this behavior was not uniform; it was qualified by significant two-way interactions with both Right-of-Way $(F(2,475.5)=4.19,p=.016)$ and Scenario $(F(6,475.0)=2.51,p=.021)$.

These significant interactions demonstrate that driver hesitance is a complex, context-sensitive judgment. The impact of the AV’s behavior was moderated by both the formal rules of the road and the physical layout of the intersection. Pairwise comparisons on the main effect of Behavior clarify the average trend: the Yield behavior elicited significantly more hesitance than both the Contingent $(p=.007)$ and Non-Yield $(p=.007)$ behaviors, which did not differ from each other (Table~\ref{tab:pairwise_combined}).

\subsubsection{Relaxation:}
The most prominent finding for relaxation was the significant and strong main effect of Behavior $(F(2,474.9)=52.15,p<.001)$. A smaller, secondary main effect was found for Scenario $(F(3,474.4)=2.72,p=.044)$, while no other factors or interactions were significant.

The lack of interactions suggests the effect of the AV’s Behavior was robust and consistent across all traffic contexts. Pairwise comparisons confirmed significant differences between all three behaviors (Table~\ref{tab:pairwise_combined}): the Contingent behavior was rated as significantly more relaxing than both Yield $(p<.001)$ and Non-Yield $(p<.001)$, and the Yield behavior was also significantly more relaxing than Non-Yield $(p<.001)$.

\subsubsection{Stress:}
Similar to relaxation, the results for stress were dominated by a significant main effect of Behavior $(F(2,474.9)=42.10,p<.001)$, and a secondary main effect of Scenario was also significant $(F(3,474.4)=2.75,p=.043)$.

\begin{figure*}[t]
\centering
\begin{minipage}{0.48\textwidth}
    \centering
    \includegraphics[width=\textwidth]{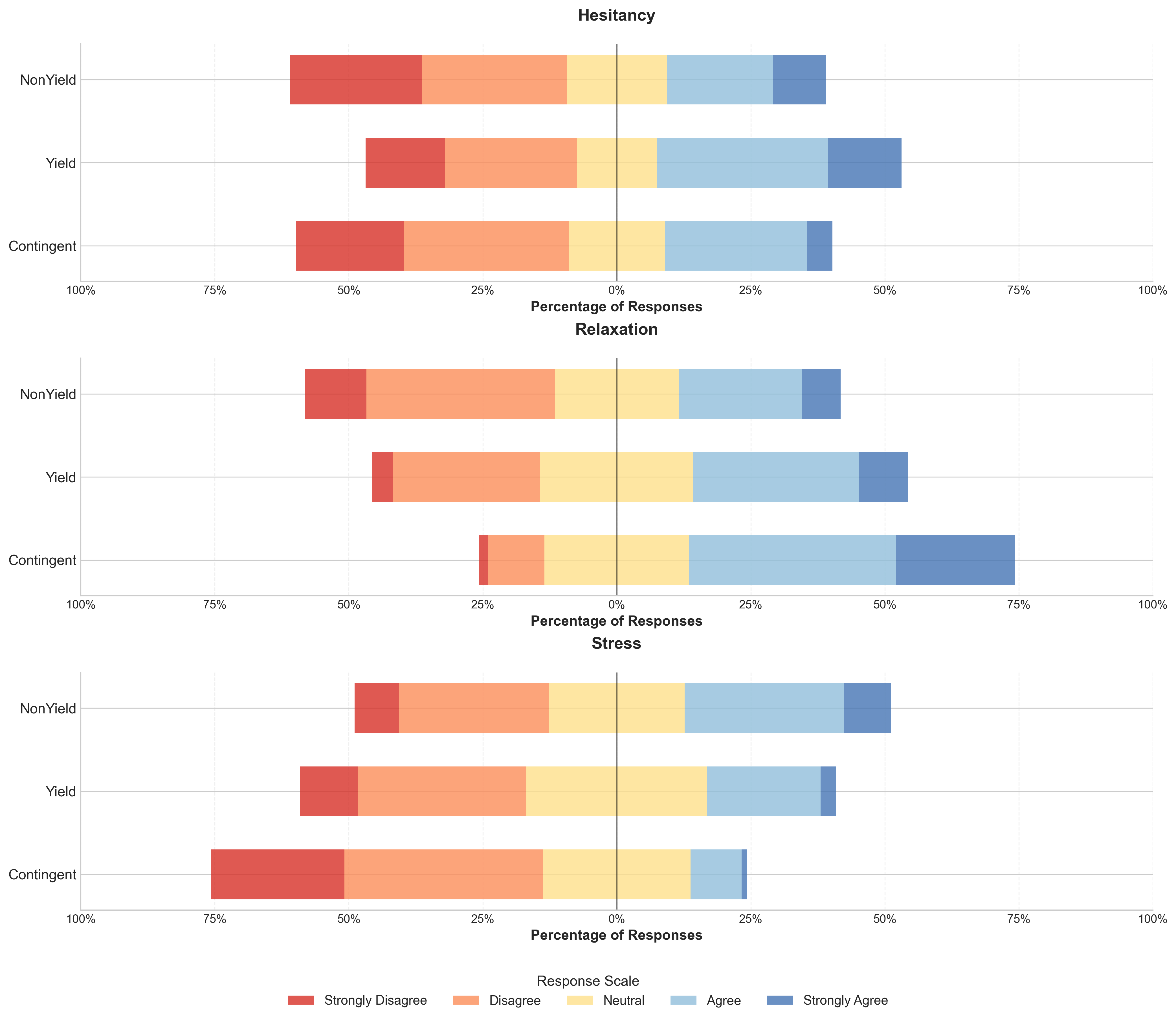}
    \caption{Bar chart showing the effect of AV Behavior on Driver's Experiences}
    \label{fig:experience-analysis}
\end{minipage}%
\hfill
\begin{minipage}{0.48\textwidth}
    \centering
       \vspace{4.5em}
    \footnotesize
        \begin{tabular}{>{\raggedright\arraybackslash}p{1.4cm}>{\raggedright\arraybackslash}p{2.4cm}cc}
    \toprule
    \textbf{Measure} & \textbf{Comparison} & \textbf{Diff.} & \textbf{Adj. p} \\
    \midrule
    \multirow{3}{*}{Hesitance} 
     & Contingent vs. Yield & -0.39 & .007 \\
     & Contingent vs. NonYield & 0.02 & 1.000 \\
     & NonYield vs. Yield & -0.41 & .007 \\
    \midrule
    \multirow{3}{*}{Relaxation} 
     & Contingent vs. Yield & 0.55 & <.001 \\
     & Contingent vs. NonYield & 0.90 & <.001 \\
     & NonYield vs. Yield & -0.35 & <.001 \\
    \midrule
    \multirow{3}{*}{Stress} 
     & Contingent vs. Yield & -0.49 & <.001 \\
     & Contingent vs. NonYield & -0.78 & <.001 \\
     & NonYield vs. Yield & 0.29 & .001 \\
    \bottomrule
    \vspace{3em}
    \end{tabular}
    \captionof{table}{Pairwise Comparisons of the AV Behaviors for the Drivers' Experiences}
    \label{tab:pairwise_combined}
\end{minipage}
\end{figure*}

This indicates that the level of stress a driver felt was primarily determined by the AV’s actions, and this effect remained consistent regardless of the specific intersection layout or right-of-way rules. Post-hoc tests for Behavior underscored this finding (Table~\ref{tab:pairwise_combined}): the Contingent behavior was significantly less stressful than both Yield $(p<.001)$ and Non-Yield $(p<.001)$, and the Yield behavior was, in turn, significantly less stressful than Non-Yield $(p=.001)$.

\subsubsection{Summary of Effects:}
When comparing across all three dependent variables (hesitance, relaxation, and stress), the results show that the \textit{Contingent Behavior} provided the best overall driver experience, most notably by evoking significantly lower stress and more relaxation compared to the other two behaviors. However, the influence of contextual factors (Right-of-way and Scenario) varied by outcome. Hesitance showed the most complex pattern, with two significant interactions: Behavior $\times$ Right-of-way and Behavior $\times$ Scenario, suggesting that people’s hesitation depended on the interplay between what the AV did and the surrounding social and situational context. In contrast, both relaxation and stress were primarily driven by Behavior, with Scenario having a smaller, independent influence. The lack of significant interactions for these outcomes highlights the consistency of the Behavior effect across all contexts. This indicates that affective responses (relaxation and stress) are more robustly determined by the AV's behavior itself, while cognitive or decision-related responses like hesitance are more sensitive to the specific situation.

\begin{table*}[t]
\centering
\caption{Type III Tests of Fixed Effects as Reported by Linear Mixed Model}
\scriptsize
\begin{tabular}{l|ccc|ccc|ccc}
\hline
 & \multicolumn{3}{c|}{Hesitation} & \multicolumn{3}{c|}{Relaxation} & \multicolumn{3}{c}{Stress} \\
\hline
Variable & $F(n_f, d_f)$ & Sig. & $\eta^2_p$ & $F(n_f, d_f)$ & Sig. & $\eta^2_p$ & $F(n_f, d_f)$ & Sig. & $\eta^2_p$ \\
\hline
Intercept & $F(1, 48.8) = 4.61$ & .037 & .086 & $F(1, 48.7) = 4.97$ & .031 & .093 & $F(1, 48.7) = 13.21$ & $<$.001 & .213 \\
Behavior & $F(2, 475.5) = 6.20$ & .002 & .025 & $F(2, 474.9) = 52.15$ & $<$.001 & .180 & $F(2, 474.9) = 42.10$ & $<$.001 & .151 \\
RightOfWay & $F(1, 48.8) = 0.02$ & .883 & $<$.001 & $F(1, 48.7) = 1.24$ & .272 & .025 & $F(1, 48.7) = 2.05$ & .158 & .040 \\
Scenario & $F(3, 474.9) = 1.46$ & .224 & .009 & $F(3, 474.4) = 2.72$ & .044 & .017 & $F(3, 474.4) = 2.75$ & .043 & .017 \\
Behavior $\times$ RightOfWay & $F(2, 475.5) = 4.19$ & .016 & .017 & $F(2, 474.9) = 2.55$ & .079 & .011 & $F(2, 474.9) = 1.54$ & .215 & .006 \\
Behavior $\times$ Scenario & $F(6, 475.0) = 2.51$ & .021 & .031 & $F(6, 474.4) = 1.84$ & .091 & .023 & $F(6, 474.4) = 2.00$ & .064 & .025 \\
RightOfWay $\times$ Scenario & $F(3, 475.0) = 0.15$ & .930 & .001 & $F(3, 474.4) = 0.13$ & .943 & .001 & $F(3, 474.4) = 0.20$ & .900 & .001 \\
Behavior $\times$ RightOfWay $\times$ Scenario & $F(6, 475.0) = 1.10$ & .364 & .014 & $F(6, 474.5) = 0.91$ & .486 & .011 & $F(6, 474.4) = 0.75$ & .612 & .009 \\
\hline
\end{tabular}
\end{table*}

\subsection{Qualitative Feedback}
At the end of the study, we conducted semi-structured interviews to gather qualitative feedback, and we also encouraged participants to verbalize their impressions of the AVs' behavior during or immediately after each interaction. Given that all AVs were visually identical and sessions were randomized, this immediate feedback was crucial for associating specific behaviors with participants' reactions. When participants exhibited strong feelings, either verbally or non-verbally, the researcher initiated brief follow-up conversations to further explore these reactions. In contrast, when we asked after the study, "Do any of the interactions stand out to you?" it resulted in general, undifferentiated responses. Although differentiating the vehicles (e.g., by changing their colors) might have made it easier for participants to compare specific interactions, this approach could have introduced new variables that might confound the results. Overall, immediate interviews and questionnaires were necessary to capture detailed responses to specific interaction situations~\cite{dey_2025_operationlize}.

\subsubsection{Impression on yielding behavior}
Out of 50 participants, 12 explicitly commented on the yielding behavior during the study. They expressed uncertainty about the AV's intentions, which created ambiguity in the interaction. As P10 stated: ``They look a little bit too hesitant, so I didn't know if I was supposed to go or not. I just waited there until they eventually went.'' Several others noted similar confusion, with P11 noting: ``They are slow so I'm not afraid of them, but the way they moved slowly with constant slow speed, I don't know if it will just slowly pass the intersection or it will wait for me. I'm hesitant and a bit confused.''

Participant P6 described the yielding behavior as ``pushy aggressive.'' When asked to elaborate on the notion, they said that ``some people in real life inch ahead, move with a constant low speed, expecting you to get done with your intersection quickly. I think it happens quite a lot but it makes me uncomfortable because their intention is very ambiguous.''

Participant P37, after interacting with the yielding behavior, told the researcher that the interaction reminded them of their past experiences with Waymo. They mentioned that ``San Francisco has many hilly intersections, so it's hard to see around the corners and stuff, as a defensive driver I usually attempt to yield to the other car, but what happened with Waymo is that we usually stopped and stared at each other for up to 30 seconds, creating an awkward moment where we both just froze there. Gestures and eye contact don't work in this case so I usually eventually still go first.''

\subsubsection{Impression on non-yielding behavior}
Of the 50 participants, 34 explicitly commented on the behavior of the non-yielding driver, with varying descriptions. Many participants expressed notions of ``aggressive'' (20 participants), ``reckless'' (15 participants), and ``incautious'' (6 participants). Some participants also noted that the non-yielding vehicle ``drives with clear intent and does not make me stressed'' (P42) [sic], and ``the fact that it's firmly deciding to go makes it easy for me to know I should yield to them, hence not a lot of mental power needed'' (P8).

When participants expressed strong negative reactions toward the non-yielding behavior, we immediately followed up with questions about familiarity, asking whether such behavior felt recognizable and why. All but two participants confirmed that the behavior felt familiar, noting that they expect to encounter such driving regularly in New York City traffic.

For example, P22 arrived in front of the intersection earlier than the Non-Yield AV in scenario \hyperref[ScenarioS2]{S2}, yet the Non-Yield AV still proceeded through the intersection. After the trial, P22 commented: ``In real life, if I see this behavior, I would feel like they are in a rush; I'll just get out of their way; and I double check to make sure there's nothing else happening in my surroundings before I move away. I'll be judgmental and think, why are you doing this? You are making everything unsafe. Did you really have to get where you're going that badly? And this behavior really reminds me of those past experiences.'' In that scenario, they rated ``very hesitant,'' ``very unrelaxed,'' and ``very stressed'' during the interaction, yet gave the full score of 10 for the familiarity question.

\subsubsection{Impression on contingent behavior}
Participants rarely spontaneously commented on the contingent behavior. In the times we probed participants to express their impressions on the contingent model after trial, participants usually didn't have much to comment on, with the common response of ``It was a comfortable interaction'' (P12, P24, P30, P33, P35) or ``it was fine'' (P8, P24, P28, P29). From this, we further deduced that the familiarity of the contingent behavior held and there was nothing conspicuous to remark on.

\subsubsection{Factors Considered for Yielding Decisions}
\label{factoryield}
During the post-study interview, when asked to describe the most important factors that allow them to decide whether to yield or not in general traffic situations, participants identified several key considerations. Arrival sequence was mentioned by 46 (92\%) participants, the turning intention of other vehicles was mentioned by 36 (72\%) participants, approach speed was identified by 30 (60\%) participants, and 4 (8\%) participants explicitly mentioned the formal traffic rule of ``yielding to the right'' when both vehicles arrive at an intersection approximately at the same time.

When reflecting on the yielding decision process, a majority of participants noted their belief that the RoW is dependent on the intended direction of turn a vehicle makes rather than the direction of approach: ``I determine RoW usually based on who's there first. If we are at the intersection at about the same time, left turn has lower priority than right turn than going straight. If there's a situation where I'm unsure about, I will typically yield.'' (P47) ``If one of us is turning left, then whoever is not turning left has the right-of-way.'' (P5) ``The vehicle not going straight should also slow down for vehicles going straight.'' (P6). It is important to note that this perception of RoW is not in accordance with the RoW rules in the US where this study took place.
\section{Discussion}
The results present our insight into how drivers respond to AV behaviors in diverse traffic contexts. Across all three measured outcomes --- hesitance, relaxation, and stress --- AV behavior consistently shaped user experience, emphasizing the critical role of motion strategy in AV design. 

\subsection{Evaluation of Hypothesis}
Our study investigated whether social familiarity of AV driving behavior affects human drivers' experiences. The results provide insights into one of our hypotheses and strong evidence supporting two of our hypotheses, demonstrating that AVs exhibiting familiar driving patterns create significantly more positive interactions across multiple experiential dimensions.

\subsubsection{H1: Hesitance}
Our first hypothesis—that a familiar contingent driving behavior would elicit less hesitance—was not fully supported, revealing a more nuanced insight. While the Contingent model did result in low hesitance, it was not statistically different from the assertive NonYield model. Instead, both of these decisive behaviors prompted significantly less hesitance than the Yield model.

This pattern of results indicates the primary factor influencing driver hesitance was not social familiarity alone, but rather the clarity and decisiveness of the AV's actions. The Yield behavior, which participants qualitatively described as ambiguous, created uncertainty and thus hesitation. In contrast, both the Contingent and NonYield behaviors projected clear, unambiguous intent, allowing human drivers to quickly predict the AV's next action and make their own decision confidently. This distinction suggests that for the cognitive task of deciding when to proceed, the clarity of an AV’s intent is more critical than its social familiarity. Indeed, the role of familiarity proves to be more influential not on cognitive judgment, but on the driver's affective experience, as the subsequent results demonstrate. 

\subsubsection{H2: Relaxation}
The result strongly supported our second hypothesis, with the Contingent model producing significantly higher relaxation scores compared to both Yield and NonYield behaviors. The medium to large effect size~\cite{cohen1988spa} indicates that a familiar contingent behavior positively impacts driver relaxation during interactions.

\subsubsection{H3: Stress}
Our third hypothesis was also strongly supported, with the Contingent model generating significantly lower stress levels compared to both non-contingent behaviors. The medium to large effect size demonstrates that familiar behavior meaningfully reduces stress in traffic interactions. The quantitative finding aligns with qualitative feedback where participants often had little remarkable to comment on about the Contingent model, and when they did, they described interactions as ``comfortable''—suggesting a natural, intuitive interaction that didn't demand cognitive effort.

The major finding from our study is that participants reported significantly less positive experiences when interacting with the non-contingent models, each of which presented a distinct set of interaction challenges. The ambiguous Yield model created uncertainty that increased driver hesitance and stress, while the assertive NonYield model, though clear in its intent, was perceived as aggressive and equally stressful. In contrast, the contingent model was remarked upon less by participants, as the contextually-adaptive algorithm trained on human driving data seemed to be perceived as more familiar and socially appropriate.

\subsection{Perception of Right-of-way}
In our study, we varied the right-of-way (RoW) condition, but the results showed that RoW as a standalone factor did not have a significant impact on drivers' overall experiences of hesitance, relaxation, or stress. However, the vast majority of our participants seemed to have a very different understanding of RoW than what is defined by US traffic law. This finding was surprising to the authors, particularly those who learned to drive in different cultures. US drivers' manuals state that at equal intersections (e.g., at four-way stops), when two vehicles arrive at the same time, the vehicle on the relative right has the RoW. However, in our study, we found that a majority of the participants felt that RoW depends on the intended direction of the vehicle. Regardless of the direction of approach, there was a noteworthy pattern: participants thought that a vehicle intending to go straight or turn right has RoW over one intending to turn left.
The consistency of this perspective suggests that drivers' interpretations of traffic situations may rely more on informal social cues than on strict awareness of rules. This aligns with studies indicating that successful traffic interaction often relies on drivers' negotiations using subtle movement cues rather than on formal adherence to rules~\cite{bin-nun_how_2022}. 

\subsection{Cultural and Contextual Factors}
A challenge of informal social cues is that they might have cultural variation which is harder to harmonize than traffic rules. There are indicators that our contingent model was well-adapted to what our study participants found familiar, but we expect that the same model would fare less well if it were deployed with users from a different driving culture, for example one which has different adherence to RoW rules. This could present interesting challenges for AV companies deploying their vehicles in new settings.

Furthermore, while the specific intersection scenario did have a statistically significant effect on stress and relaxation, its practical impact was minor compared to the influence of the AV's behavior. Our analysis shows the effect size of Behavior ($\eta^2_p \approx .15-.18$) was nearly an order of magnitude larger than that of Scenario ($\eta^2_p = .017$). This finding strongly reinforces that the quality of the dynamic interaction, rather than the static spatial arrangement, is the primary determinant of a driver’s subjective experience. This suggests that AV designers should focus on developing context-sensitive interaction models rather than scenario-specific rules.

\subsection{Implications for AV Design}
Our study shows that human drivers respond more favorably to the data-driven contingent AV algorithm, which creates a significantly more positive affective experience by reducing stress while still fostering a clear, decisive interaction. This suggests that AVs designed with a more human-like contingency can foster a sense of social familiarity and mutual trust between AVs and human drivers, thereby boosting AV acceptance, as trust is the most identified determinant~\cite{zhang_what_2021, haghzare_older_2021}. Moreover, the capability for AVs to predict human driver actions and adjust accordingly can also enhance the overall traffic flow by reducing hesitance and conflicts at intersections, which is a crucial aspect of integrating AVs into current traffic systems.

Our study findings advocate a paradigm shift from the modularized hand-engineered approach to an end-to-end data-driven approach to autonomous driving systems~\cite{Chen2023End-to-End}. Systems should be able to adjust dynamically based on situation awareness rather than relying solely on fixed rule sets. AVs should also demonstrate effective behavioral cues that make their intentions clear to other road users, including human drivers and AVs.

This shift holds significant potential for handling complex urban scenarios and edge cases. It is imperative to address the associated challenges of interpretability and safety. The "black box" nature of deep learning models can make them difficult to understand, raising concerns about their lack of transparency and reliability. By integrating methods that enhance interpretability and align machine behavior with human expectations, autonomous systems will not only be efficient and adaptable but also trustworthy and socially acceptable ~\cite{Chen2023End-to-End}.

\subsection{Limitations}

A primary limitation of this study is its reliance on a Virtual Reality (VR) driving simulator. We chose VR to create a controlled environment, allowing for the isolation of key variables, and also mitigating potential physical risk to participants. However, this control comes at the cost of ecological validity, as a simulation cannot fully replicate the dynamic conditions of real-world traffic. The driving dynamics of the simulated vehicle differ from a physical car, and driver decision-making may be altered by the absence of real-world consequences. These differences could limit the direct generalizability of our findings to actual traffic interactions, which are shaped by a wider range of environmental and social pressures. To help mitigate these factors, participants completed a practice drive to familiarize themselves with the simulation environment before the study began.

Additionally, the findings may be shaped by the traffic norms and social behaviors in the specific locale where the study was conducted. Some of the insights may not translate to regions with different driving cultures or regulatory environments. Although not a central limitation, this highlights the importance of taking cultural context into account in traffic interaction research. Work is underway to explore and validate our findings in diverse settings. 

\subsection{Artifacts Available}

The following supplementary documents are available on the ACM DL in association with this publication. We encourage other researchers to make use of these resources and to adopt our study protocol and software:

\begin{itemize}
    \item \textbf{Interaction Data}: The simulator records comprehensive data including vehicle motion, head orientation, hand gestures, steering inputs, pedal usage, turn signals, and other driver actions. Though we did not use these data for this paper, we will use them for future publications, and the dataset is made publicly available as supplementary material.
    \item \textbf{Questionnaire Data}: The questionnaire data for analysis is made available as supplemental material.
    \item \textbf{Study Session Sample}: A video showing the procedure of the study session.
    \item \textbf{Simulation Environment}: The deployment on Driver-AV interaction in the StrangeLand simulator is made public as GitHub Repository.
    \item \textbf{AV Training}: The training and deployment code for the AV is made public as GitHub Repository.
    
\end{itemize}
\section{Conclusion}
This paper presents a mixed-method, VR-based experiment investigating human driver experiences with autonomous vehicles. We compared driver reactions to a familiar contingent AV, which responds dynamically based on social norms, against non-contingent AVs that follow rigid, predetermined patterns. Empirical evidence shows that a familiar, contingent AV behavior has an overall positive effect on drivers' experiences by simultaneously reducing stress and promoting relaxation while fostering the clear, decisive interactions that reduce driver hesitance. This positive impact is further substantiated by subjective feedback from participant interviews. This highlights the need for AV algorithms to account for the implicit behavior of other drivers. Incorporating social familiarity is crucial for integrating AVs seamlessly into traffic as socially aware agents.
\begin{acks}
This research was made possible by NSF research grant IIS-2107111: Cultural Differences in Driving Interaction. We would like to thank the facilities staff of Cornell University for their support. We are also grateful to Dr. David Goedicke for his help in developing the system used in the \href{https://github.com/Strange-Land}{StrangeLand Simulator}.

\end{acks}

\bibliographystyle{ACM-Reference-Format}
\bibliography{zotero,references}

\clearpage
\appendix
\onecolumn

\section{Experimental Trial Structure}
\label{app:sessions}
The table below lists the 12 unique trial types presented to each participant group. The presentation order of these trials was randomized for each participant during the study.

\begin{table*}[htp]
\centering
\begin{tabular}{ccc|ccc}
\hline
\multicolumn{3}{c|}{Group A Sessions (AV has RoW)} & \multicolumn{3}{c}{Group B Sessions (Driver has RoW)} \\
\hline
Session Number & Scenario & Behavior & Session Number & Scenario & Behavior \\
\hline
1 & S1 & Contingent & 1 & S1 & Contingent \\
2 & S1 & Yield & 2 & S1 & Yield \\
3 & S1 & NoYield & 3 & S1 & NoYield \\
4 & S2 & Contingent & 4 & S2 & Contingent \\
5 & S2 & Yield & 5 & S2 & Yield \\
6 & S2 & NoYield & 6 & S2 & NoYield \\
7 & S3 & Contingent & 7 & S3 & Contingent \\
8 & S3 & Yield & 8 & S3 & Yield \\
9 & S3 & NoYield & 9 & S3 & NoYield \\
10 & S4 & Contingent & 10 & S4 & Contingent \\
11 & S4 & Yield & 11 & S4 & Yield \\
12 & S4 & NoYield & 12 & S4 & NoYield \\
\hline
\end{tabular}
\caption{Participant Sessions for Groups A and B}
\label{tab:participant_sessions}
\end{table*}

\section{Post-Interaction Questionnaire Items}
\begin{table*}[htp]
\centering
\begin{tabular}{p{7cm}p{7cm}}
\hline
\textbf{Question} & \textbf{Response Options} \\
\hline
\textbf{Variable Hesitance:} How hesitant were you about your course of action in the interaction space? & 
Very hesitant, Hesitant, Neutral, Not Hesitant, Not Hesitant at All \\
\hline
\textbf{Variable Relaxation:} How relaxed did you feel during the interaction with the AV? & 
Very un-relaxed, Un-relaxed, Neutral, Relaxed, Very relaxed \\
\hline
\textbf{Variable Stress:} To what extent did you find this interaction stressful? & 
Very stressful, Stressful, Neutral, Not stressful, Not at all stressful \\
\hline
\textbf{Variable Familiarity:} On a scale of 1 to 10, how would you rate the other vehicle's behavior in this experience compared to how you normally see in real-world traffic? (With 0 being ``nothing like the real world'' and 10 being ``exactly like the real world'') & 
Slider scale from 0 to 10 \\
\hline
\end{tabular}
\caption{Post-Interaction Questionnaire Items}
\label{tab:questionnaire_items}
\end{table*}

\end{document}